\documentclass{article}
\usepackage{spconf,amsmath,graphicx}
\usepackage{amssymb}
\usepackage{amsmath}
\usepackage{booktabs} 
\usepackage{xcolor}
\usepackage[normalem]{ulem}
\def\bX{\boldsymbol{X}}
\def\bx{\boldsymbol{x}}
\def\bc{\boldsymbol{c}}
\def\bs{\boldsymbol{s}}
\def\mos{{o-MOS }}

\usepackage{hyperref}
\newcommand{\para}[1]{\vspace{5pt}\noindent \textbf{#1} \, }

\title{Training Robust Zero-Shot Voice Conversion Models with Self-supervised Features}
\name{Trung Dang$^{\star}$\thanks{Work performed while Trung Dang was an intern at Microsoft.}, Dung Tran$^{\dagger}$, Peter Chin$^{\star}$, Kazuhito Koishida$^{\dagger}$}
\address{$^{\star}$ Boston University \quad
      $^{\dagger}$ Microsoft Corporation}
\begin{document}
%
\maketitle
\begin{abstract}

Unsupervised Zero-Shot Voice Conversion (VC) aims to modify the speaker characteristic of an utterance to match an unseen target speaker without relying on parallel training data. Recently, self-supervised learning of speech representation has been shown to produce useful linguistic units without using transcripts, which can be directly passed to a VC model. 
In this paper, we showed that high-quality audio samples can be achieved by using a length resampling decoder, which enables the VC model to work in conjunction with different linguistic feature extractors and vocoders without requiring them to operate on the same sequence length. We showed that our method can outperform many baselines on the VCTK dataset. Without modifying the architecture, we further demonstrated that a) using pairs of different audio segments from the same speaker, b) adding a cycle consistency loss, and c) adding a speaker classification loss can help to learn a better speaker embedding. Our model trained on LibriTTS using these techniques achieves the best performance, producing audio samples transferred well to the target speaker's voice, while preserving the linguistic content that is comparable with actual human utterances in terms of Character Error Rate.


\end{abstract}
\section{Introduction}
\label{sec:intro}

The task of alternating the voice characteristic of someone's speech as if spoken by another target speaker has always been an interesting branch of style transfer learning, with applications ranging from entertainment to protection of the speaker's identity. Traditional voice conversion systems require training on parallel data and/or audio transcription, and only target one or several voices of seen speakers. However, recent architectures have been able to perform unseen-to-unseen, or zero-shot, voice conversion in a completely unsupervised manner. Some previously proposed architectures include conditional variational autoencoders (CVAE) \cite{hsu2016voice,saito2018non}, generative adversarial networks (GAN) \cite{kameoka2018stargan,zhang2020gazev,nguyen2021nvc}, or autoencoders with feature disentanglement \cite{qian2019autovc,chou2019one}. 

Recently, works on self-supervised representation learning (SSL) in the speech domain have been able to produce speech units by solving a contrastive task \cite{oord2018representation,baevski2020wav2vec}. These speech features can be obtained without training on labeled data, which perfectly suits the voice conversion task, where it is important to learn phonetic representations, but not necessary to map them to actual phonemes. Despite producing competitive results compared to traditional approaches, two challenges remain in the SSL approach. First, with the content embedding fixed, VC models with a frame-by-frame decoder become less flexible. While the speech units can be efficiently obtained from speech features of a lower resolution, vocoder benefits from a smaller window size and a higher sampling rate. Moreover, the speaker embedding needs to be more feature-rich, both well representing the speaker's voice characteristic and distinguishing itself from embeddings by other speakers.

To address the aforementioned problems, in this paper, we propose a VC model with a length resampling decoder that works with length-incompatible input and output features, allowing the self-supervised acoustic model and the vocoder to be trained on their task-optimal speech features. We also demonstrate that training on pairs of segments from the same utterance and adding a cycle-consistency loss can help to learn a lower variant speaker embedding, which leads to higher audio quality. Moreover, we show that training a speaker encoder jointly with a speaker classification loss produces utterances with better-transferred voice characteristics. We combine these training techniques and demonstrate performance improvement in terms of objective and subjective evaluation, on the VCTK dataset and the LibriTTS dataset.

\section{Related Works}
\label{sec:related_works}

In this section, we revisit recent works on unsupervised models that can perform zero-shot voice conversion. 

\para{CVAE and GAN-based} Conditional Variational Autoencoders (CVAEs)
have been successfully applied to the VC task \cite{hsu2016voice,saito2018non}. While it allows models to be trained without parallel data, its quality is usually poor due to over smoothed decoder outputs. On the other hand, Generative Adversarial Networks (GANs), with their distribution matching property, are shown to produce better audio quality \cite{kameoka2018stargan,zhang2020gazev,nguyen2021nvc}, though they are notoriously hard to train. 

\para{Autoencoders with Feature Disentanglement} Voice conversion can ideally be achieved by pulling out speaker-related information from an utterance and plugging in information from another speaker. For this to be feasible, we desire to disentangle speaker embedding (global, time-invariant) with content embedding (local, time-variant). This can be achieved by precise embedding size tuning \cite{qian2019autovc}, adaptive instance normalization \cite{chou2019one,chen2021again}, or vector quantization \cite{wu2020one}. These frameworks have been shown to have better performance compared to CVAE and GAN-based counterparts. The disentanglement can also be specifically achieved with $\beta$-VAE-style loss \cite{luong2021many} or minimizing mutual information \cite{yuan2021improving}. However, some architectures rely on an information bottleneck that can affect the audio quality. These methods also do not explicitly leverage linguistic content in the utterance, thus sometimes produce unclear or distorted samples.

\para{Using Self-supervised Acoustic Features} Early works have utilized speaker-independent acoustic features from a pre-trained ASR for the VC task \cite{lu19_interspeech}; however, this requires an extensive amount of labeled data for training the ASR. Recent works show that self-supervised speech representations can be used instead \cite{lin2021fragmentvc,lin2021s2vc,choi2021neural}. Since the content embedding is fixed as an input to the VC model, it is important to learn a good speaker embedding. Our work aims to design an architecture that can be efficiently used in conjunction with existing acoustic feature extractors and vocoders. We also propose techniques to produce speaker embeddings that are more useful to the task.

\section{Methodology}


First, let us introduce the conventional framework. Let $\bX_u$ and $\bX_v$ be the source and target audio samples. Let $\bx_u$ and $\bx_v$ denote the speech features (e.g. mel filter banks or MFCCs). We desire to produce speech features of the same length as $\bx_u$ (we denote the length as $d_{\text{out}}$). The content embedding of the source utterance $\bc_u$ is obtained from the raw audio: $\bc_u=E_\text{c}(\bX_u)\in\mathbb{R}^{d_{\text{inp}}\times d_c}$,  while the speaker embedding of the target utterance $\bs_v$ is obtained from the speech features: $\bs_v=E_\text{s}(\bx_v)\in\mathbb{R}^{d_s}$, where $E_c$ and $E_s$ are the content and the speaker encoder, respectively, $d_{\text{inp}}$ is the length of the content embedding, and $d_s$, $d_c$ are the embedding sizes. A decoder $D(\bc,\bs): \mathbb{R}^{d_s}\times\mathbb{R}^{d_{\text{inp}}\times d_c}\mapsto\mathbb{R}^{d_{\text{out}}\times d}$ produces the $d$-dimensional speech features for audio generation from $\bc_u$ and $\bs_v$. During training, since parallel data is not available, the samples used for the source and the target utterance are identical, thus $\bX_u=\bX_v=\bX$, $\bc_u=\bc_v=\bc$, and $\bs_u=\bs_v=\bs$. The self-reconstruction loss is defined as $\mathcal{L}_{\text{self-same}}=\|\bx-D(\bc, \bs)\|_2$

Within the above framework, in subsection \ref{subsec:resampling}, we propose a length resampling decoder, which enables the VC model to operate on inputs and outputs of different (but proportional) length (i.e., $d_\text{in}\ne d_\text{out}$). We also cover, in the \ref{subsec:self-reconstruction} and \ref{subsec:consistency} subsections, three proposed techniques aiming to train a better speaker encoder.

\subsection{Length resampling decoder}\label{subsec:resampling}

In the conventional architecture, the decoder receives the content embedding and produces speech features, which are used to generate audio, either through a reconstruction algorithm or a neural-based vocoder. A problem arising when generating speech features for audio generation from self-supervised acoustic features is the length mismatch. While vocoder may benefit from a higher sampling rate and lower window size, speech representation can be more effectively obtained from features with a longer window size, or directly from raw audio. Though the sequence length of the input $d_{\text{inp}}$ and the output $d_{\text{out}}$ of the decoder are different, they are proportional. To map sequences of fixed length ratio, we propose using a sequence of upsampling layers to produce high-resolution features and perform a pooling average at the end. For example, if the content encoder receives 16kHz audio samples and produces features with 320x reduction, and the vocoder produces 24kHz audio samples from speech features with a window size of 300, the $d_{\text{inp}}:d_{\text{out}}$ ratio is 5:8. We can upsample the decoder input by a factor of 8, and perform an average pooling of stride 5. This allows the self-supervised acoustic model and the vocoder to be trained independently on their task-optimal preprocessing configurations.

\subsection{Self-reconstructing from pairs of utterances}\label{subsec:self-reconstruction}

A majority of prior works train the model with $\bx_u=\bx_v=\bx$. While $\bx$ should be self-reconstructed from its own features, i.e. $\bx=D(E_c(\bx), E_s(\bx))$, for feature disentanglement, we find this training schema different from the inference setting when $\bx_u\ne\bx_v$. Moreover, speaker embeddings could have access to content information used for self-reconstruction, which is undesirable. We modify the training process by allowing the model to be trained on pairs of audio segments from the same speaker. Specifically, the loss becomes
\begin{equation}\mathcal{L}_{\text{self-diff}}=\|\bx_u-D(\bc_u, \bs_v))\|_2+\|\bx_v-D(\bc_v, \bs_u))\|_2\end{equation}

Since for the same speaker, the voice characteristic also varies with the spoken context, we only sample segments from the same utterance.

\subsection{Enforcing consistency and separability for speaker embeddings}\label{subsec:consistency}

While the speaker embedding can be obtained from a speaker verification/classification model pre-trained on a large corpus with speaker ids \cite{qian2019autovc}, it has been shown that a jointly trained speaker embedding without using the speaker identity is more suitable for VC tasks \cite{huang2021far}. 

To encourage the speaker embedding to be invariant for the same speaker, we add a loss term for cycle-consistency. Let $\bx'_u$ and $\bx'_v$ be the speech features produced by the decoder (i.e. $\bx'_u=D(\bc_u, \bs_v)$ and $\bx'_v=D(\bc_v,\bs_u)$). The cycle-consistency loss is defined as
\begin{equation}\mathcal{L}_{\text{cycle}}=\|\bs_{u}-\bs_{v}\|_2+\|\bs_{u}-E_s(\bx'_u)\|_2+\|\bs_{v}-E_s(\bx'_v)\|_2\end{equation}

The cycle-consistency loss pushes speaker embeddings of segments of the same speaker closer. Speaker embeddings of utterances by different speakers should remain distinguishable since they must share mutual information with the corresponding utterances for the self-reconstruction to be feasible. However, minimizing this loss may also make the speaker embeddings of different speakers less separable. We resolve this by additionally utilizing \textit{speaker id} information. A projection head is added on top of the speaker embedding for classification. Let $y$ represent the one-hot embedding of the speaker of sampled utterances, and $\mathcal{L}_{\text{xent}}$ be the cross-entropy loss. The speaker id loss is as follows
\begin{equation}\mathcal{L}_{\text{speaker}}= \mathcal{L}_{\text{xent}}(\text{Proj}(\bs_{u}),y)+\mathcal{L}_{\text{xent}}(\text{Proj}(\bs_{v}), y)\end{equation}

\begin{figure}[t]
    \centering
    \includegraphics[width=0.5\textwidth]{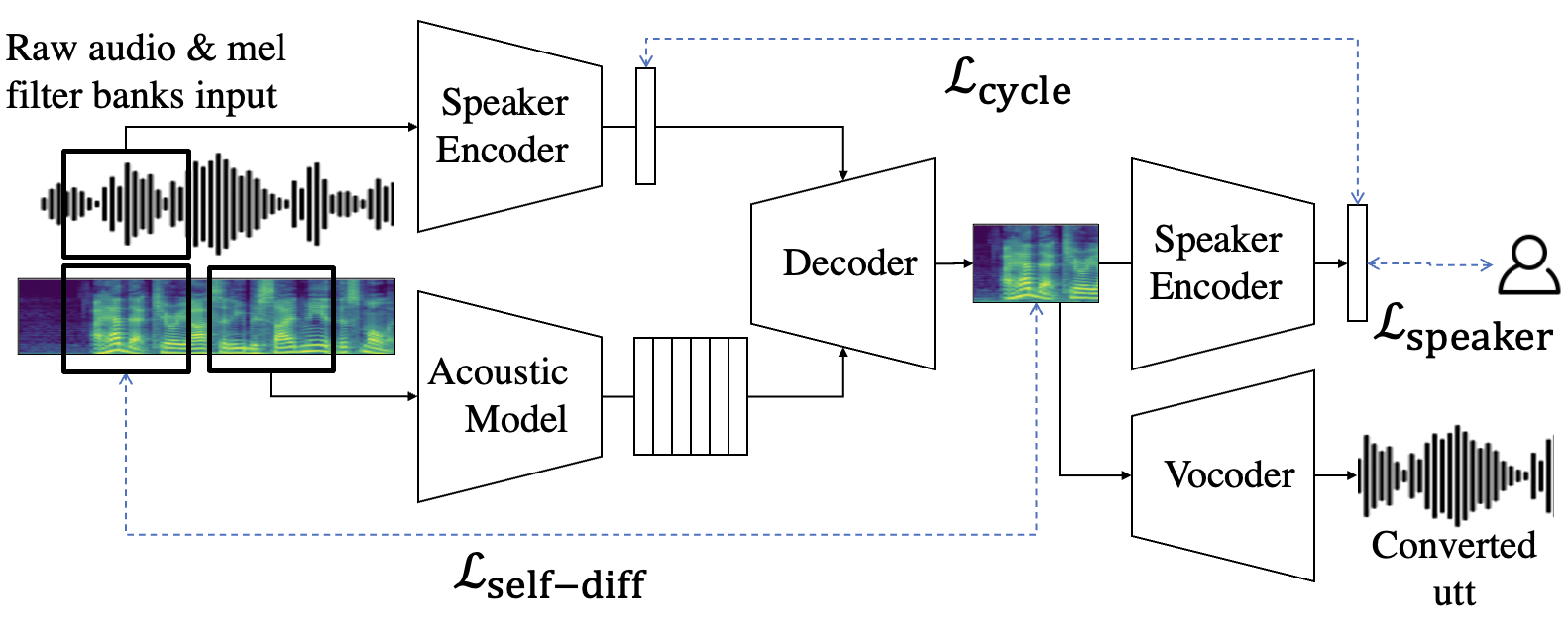}
    \caption{Our proposed methods during the training stage. Two segments sampled from the same utterance are used to extract the speaker embedding and the acoustic features. The decoder fuses the speaker embedding into each decoding frame, also resamples it to match the length of features required for the vocoder. A cycle-consistency loss $\mathcal{L}_{\text{cycle}}$ and a speaker classification loss $\mathcal{L}_{\text{speaker}}$ are optionally used.}
    \label{fig:method}
\end{figure}

Figure \ref{fig:method} provides an illustration of our method. 

\section{Experiments}
\label{sec:experiments}

\subsection{Experimental Setup}

\para{Baselines} We compare our method with other self-supervised architectures that can perform zero-shot voice conversion. We use publicly available checkpoints of AutoVC\footnote{Checkpoints available at \url{https://github.com/auspicious3000/autovc}} and FragmentVC\footnote{Checkpoints available at \url{https://github.com/yistLin/FragmentVC}} trained on VCTK. We also follow instructions to train the AdaIN-VC baseline.

\para{Datasets} We adopt the VCTK dataset \cite{Veaux2016SUPERSEDEDC}, which is commonly used in training VC models. Since the train/test splits of VCTK are inconsistent for baselines, we need to resample testing data. Specifically, we randomly sample 10 speakers from the whole dataset, from which we further sample 100 pairs of utterances for testing. We train our model only on the remaining 100 speakers\footnote{Test utterances could have been used for the baseline checkpoints}. 
Note that some baselines use additional datasets, for example, Auto-VC uses a speaker embedding model pre-trained on VoxCeleb, or FragmentVC uses a wav2vec feature extractor pre-trained on LibriSpeech. Our results also use LibriSpeech/LibriTTS for the acoustic feature extractor and the vocoder. For a more integrated setting, we also set up a LibriTTS benchmark in which the whole VC pipeline is trained on LibriTTS\footnote{LibriSpeech can be alternatively used.}. Using LibriTTS enables us to test models on utterances recorded in a more diversified and realistic setting. We sample 1,000 pairs of source/target utterances with different speakers from the test-clean set of LibriTTS for evaluation. 
We also set up a validation set from the dev-clean subset of LibriTTS and report the results with the lowest self-reconstruction loss. 

\begin{table*}[t]
\caption{Objective (SVA, \mos, CER) and subjective scores (NAT, SIM) of the AdaIN-VC baseline and our methods on LibriTTS dataset. For reference, we also report scores from human recorded utterances when available.}
\label{tab:libritts}
\centering
\begin{sc}
\begin{tabular}{l c cccccc c cccccc}
\toprule
& & \multicolumn{6}{c}{LibriTTS 100hr} &  & \multicolumn{6}{c}{LibriTTS 460hr} \\
\cmidrule{3-8} \cmidrule{10-15}
 & & SVA & \mos & CER & & NAT & SIM & & SVA & \mos & CER & & NAT & SIM \\
\cmidrule{1-1}\cmidrule{3-5}\cmidrule{7-8}\cmidrule{10-12}\cmidrule{14-15}
Original & & N/A & 3.26 & 4.15 & & 4.56 & N/A & & N/A & 3.26 & 4.15 & & 4.56 & N/A \\
AdaIN-VC \cite{chou2019one} & & 0.71 & 2.79 & 14.87 & & 2.50 & 1.85 & & 0.86 & 2.92 & 15.48 & & 2.71 & 1.88 \\
\cmidrule{1-1}\cmidrule{3-5}\cmidrule{7-8}\cmidrule{10-12}\cmidrule{14-15}
$\mathcal{L}_{\text{self-same}}$ & & \textbf{0.91} & 2.84 & 8.08 & & 3.02 & \textbf{2.11} & & \textbf{0.97} & \textbf{2.94} & 7.79 & & 3.39 & \textbf{2.34}\\
$\mathcal{L}_{\text{self-diff}}$ & & 0.82 & 2.76 & 7.20 & & 2.71 & 1.85 & & 0.92 & 2.93 & 6.09 & & 3.31 & 2.01 \\
$\quad+\mathcal{L}_{\text{cycle}}$ & & 0.83 & 2.76 & \textbf{6.94} & & 2.91 & 1.86 & & 0.92 & 2.90 & 5.90 & & 3.51 & 2.08 \\
$\quad+\mathcal{L}_{\text{speaker}}$ & & 0.88 & 2.87 & 7.55 & & 3.16 & 1.95 & & 0.95 & 2.90 & \textbf{4.15} & & \textbf{3.53} & 2.15 \\
$\quad+$both & & \textbf{0.91} & \textbf{2.90} & 6.95 & & \textbf{3.41} & 1.97 & & 0.95 & 2.91 & 4.25 & & 3.31 & 2.29 \\
\bottomrule
\end{tabular}
\end{sc}
\end{table*}

\begin{table}[t]
\setlength\tabcolsep{4pt}
\caption{Objective (SVA, \mos, CER) and subjective scores (NAT, SIM) on the VCTK dataset. Utterances produced by our method appears to preserve well the linguistic content, have good perceptual quality, and be similar to the target voice}
\label{tab:vctk}
\centering
\begin{sc}
\begin{tabular}{l c ccc c cc}
\toprule
& & SVA & \mos & CER & & NAT & SIM \\ 
\cmidrule{1-1}\cmidrule{3-5}\cmidrule{7-8}
AutoVC \cite{qian2019autovc} & & 0.21 & \textbf{3.08} & 68.25 & & 2.31 & 1.68 \\  
AdaIN-VC \cite{chou2019one} & & 0.46 & 2.80 & 24.61 & & 2.58 & 1.89 \\
FragVC\footnote{Requires concatenation of 5 utterances for the target} \cite{lin2021fragmentvc} & & \textbf{0.71}  & 2.83 & 32.71 & & 3.24 & 2.08 \\ 
\cmidrule{1-1}\cmidrule{3-5}\cmidrule{7-8}
$\mathcal{L}_\text{self-same}$ & & 0.64 & 2.88 & 9.79 & & 3.53 & \textbf{2.23} \\
$\mathcal{L}_\text{self-diff}$ & & 0.54 & 2.91 & \textbf{6.15} & & \textbf{3.66} & 2.12 \\
\bottomrule
\end{tabular}
\end{sc}
\end{table}

\para{Model Architecture \& Training Details} We use the wav2vec2 checkpoint released by the authors for the self-supervised acoustic model, which is trained on 960 hours of audio from the LibriSpeech dataset. For the vocoder, we use the official checkpoint released with ParallelWaveGAN \cite{yamamoto2020parallel}, both available for the VCTK and the LibriTTS dataset. We do not fine-tune these models, but only train our VC models with input and output following these pre-trained models' data preprocessing recipes. Specifically, the 256-dimensional projected states of wav2vec2 are given to the models (this is different from FragmentVC where the last hidden states of 768-dimensional vectors are used). The vocoder receives decoded speech features with a window size of 300 and generates 24kHz audio. Our speaker encoder consists of 12 stacked 1D convolutional neural network (CNN) layers with and kernel size 5, followed by 12 fully-connected layers, all with a hidden size of 128 and residual connections. Our decoder consists of 12 CNN layers with PixelShuffle layers \cite{shi2016real} for 8x upsampling. The output is then down-sampled by a factor of 5. Our VC models are trained on (1) 100 speakers from the VCTK dataset (2) 100 hours and (3) 100+360 hours of clean speech from the LibriTTS dataset \cite{zen2019libritts}. All utterances are randomly cropped or padded to 2-second audio segments. We train our models with 100k steps for VCTK and LibriTTS-100hr, and 500k steps for LibriTTS-460hr, with the batch size of 64 and a constant learning rate of $5\times 10^{-4}$.

\subsection{Evaluation}

\paragraph*{Objective Evaluation} We evaluate converted utterances of each model on three tasks: 
(1) Speaker Verification Accuracy (SVA): We use Resemble.ai, a third-party speaker verification service to compute the similarity scores between the target and converted utterances. The conversion is considered successful if the score is greater than a threshold, which is decided by the equal error rate (EER) over 100/1000 samples from VCTK/LibriTTS 
(2) Character Error Rate (CER): We use Speech to Text in Microsoft Azure to obtain the transcript of converted utterances. We compute CER scores with regards to the original transcripts to evaluate the intelligibility of converted utterances. (3) Objective Mean Opinion Score (o-MOS): We train a MOSNet \cite{lo2019mosnet} on the human evaluation results of the Voice Conversion Challenge (VCC) 2018, to provide an objective metric that correlates to human's evaluation. 

\para{Subjective Evaluation} Following prior works, we provide the naturalness score based on the Mean Opinion Score (MOS) and the speaker similarity score.
(1) Speaker Similarity (SIM): Each subject is asked to listen to the target and converted utterances, and give to score from 1 to 3, which corresponds to \textit{not similar} / \textit{unsure} / \textit{similar}, regarding how confident they would consider these two utterances to be spoken by the same speaker. (2) Naturalness (NAT): Each subject is asked to listen to the converted utterance and give a score from 1 to 5 regarding how natural it sounds. Subjective evaluation is performed on 100 utterances for each model. Each audio sample is rated by 7 subjects from a pool of 1000+ raters.

\subsection{Results}

\paragraph*{Using VCTK} Table \ref{tab:vctk} shows evaluation results on baselines and our proposed methods using only the self-reconstruction loss. For baseline methods, the CER scores are substantially higher than the actual human speech, and far above the scores of utterances generated with our models. On subjective evaluation, our models outperform the best baseline FragmentVC on NAT score by 0.42 (relatively 13\%) for $\mathcal{L}_{\text{self-diff}}$, and on NAT score by 0.15 (relatively 7.2\%) for $\mathcal{L}_{\text{self-same}}$. 

\para{Using LibriTTS} The results for LibriTTS are presented in table \ref{tab:libritts}. Since pre-trained checkpoints are not available, we only train and compare our results with AdaIN-VC. It can be observed that training with $\mathcal{L}_{\text{self-diff}}$ and training with $\mathcal{L}_{\text{cycle}}$ sometimes improve the audio quality, but degrade the speaker similarity score. This suggests that though these techniques can reduce the variance for speaker embeddings from the same speaker, they also make speaker embeddings from different speakers less separable. Our model trained with the speaker classification loss produces a more balanced result between audio quality and speaker similarity. When training on LibriTTS-460hr, our models achieve comparable CER scores with human recorded voices. We encourage our readers to listen to these audio samples\footnote{\url{https://trungd.github.io/ssl_vc/index.html}}.

\section{Conclusion}

We demonstrated that high-quality voice conversion systems can be achieved by training on self-supervised speech units with our proposed architectures and training strategies. 
Our models produce good perceptual quality utterances with similar voice characteristic as the target speaker, while maintaining the linguistic content of the source utterance.

\newpage

\bibliographystyle{ieeetr}
\bibliography{strings,refs}

\end{document}